\documentclass[aps,prb,superscriptaddress,showpacs,preprintnumbers,twocolumn,amsmath,floatfix]{revtex4}
\usepackage{amsmath, amsfonts, amssymb, mathrsfs}
\usepackage{mathptmx}
\DeclareMathAlphabet{\mathcal}{OMS}{cmsy}{m}{n}
\usepackage{graphicx}
\usepackage{fancybox}
\usepackage{gensymb}
\usepackage{pifont}
\usepackage{bm}
\usepackage[dvipsnames]{color}
\usepackage{url}
\usepackage{multirow}
\usepackage{array}
\usepackage[colorlinks, linkcolor=blue, citecolor=black, urlcolor=blue]{hyperref}

\begin{document}


\title{2D clathrate graphene in minimum egg-tray-shape: An \textit{ab initio} study}




\author{Guohui Zheng}
\email{ghzheng@gzu.edu.cn}
\affiliation{School of Physics, Guizhou University, Guiyang 550025, P. R. China}

\author{Xiaosi Qi}
\affiliation{School of Physics, Guizhou University, Guiyang 550025, P. R. China}

\date{\today}

\begin{abstract}
  The thriving area of synthetic carbon allotropes witnesses
  theoretic proposals and experimental syntheses of many new two-dimensional ultrathin structures,
  which are often achieved by careful arrangement of non-hexagon $\mathrm{sp^2}$ defects in graphene.
  Here, we introduce pyramid $\mathrm{sp^3}$ hybridization into $\mathrm{sp^2}$ network
  and propose a new carbon polymorph with clathrate pattern and with minimum egg-tray shape
  (termed as clathrate graphene).
  Eight symmetrically equivalent $\mathrm{sp^2}$ carbon atoms and two symmetrically equivalent $\mathrm{sp^2}$ carbon atoms
  in its tetragonal primitive unit cell form two perpendicularly oriented rectangles and four bridging hexagons.
  Though deformed bond lengths and bond angles, the planar geometry of both tetrarings and hexagons are retained.
  High percentage and small deformation of hexagons make this metastable $\mathrm{sp^2-sp^3}$ allotrope
  comparable with pure $\mathrm{sp^2}$ $T$-graphene and penta-graphene in energetics.
  Exhaustive \textit{ab initio} calculations confirm its dynamical and elastic stabilities,
  reveal its semiconducting nature with an indirect band gap of 0.90 eV for unstressed sample,
  and suggest a giant strain tuning effect which endows versatile electronic properties ranging from metallic to semiconducting.
  Furthermore, we observe multiple von Hove singularities near the Fermi energy.
  These salient properties may imply potential nanoelectronic applications.
  These findings help understand structure-property relationship for two-dimensional carbon allotropes,
  and help search new carbon polymorphs.
\end{abstract}

\maketitle

Carbon, with electron configuration of $\mathrm{[He]2s^22p^2}$, has most redundant and versatile compound forms
ranging from organic to inorganic and with various hybridization forms and diverse structure complexities.
After thousands year of usage of diamond and graphite,
carbon science regains its vitality and branches out to the era of synthetic carbon allotropes~\cite{Hirsch},
following the successive experimental triumph of zero-dimensional (0D) fullerene~\cite{Iijima_fullerene},
one-dimensional (1D) carbon nanotube (CNT)~\cite{Iijima_CNT}
and two-dimensional (2D) graphene~\cite{Novoselov_graphene, Castro_graphene}.
The phenomenological interests in synthetic carbon allotropes originate from
their novel electronic and functional properties,
such as superhardness in monoclinic M-carbon~\cite{Li_Mcarbon},
orbital-frustration-induced metal-insulator transition in carbon kagome lattice~\cite{Chen_CKL},
and topological node-line semi-metallicity in body-centered tetragonal $\mathrm{C_{16}}$~\cite{Cheng_BCTC16}.
In synthetic carbon science,
synergy effect between theories and experiments witnesses the birth of many new forms of carbon allotropes,
such as three-dimensional (3D) T-carbon~\cite{Sheng_Tcarbon,Zhang_TcarbonNW,Xu_Tcarbon}
and 2D graphyne and graphdiyne~\cite{Li_graphdiyne}.

In 2D aspect, the superstar is the one-atomic thick graphene with honeycomb lattice,
which possesses many salient properties,
such as massless Dirac fermion and ultrahigh electronic and phononic conductivity
~\cite{Novoselov_graphene, Castro_graphene,Balandin_graphene}.
Moreover, graphene is building material for many carbon materials of various dimensionalities,
and can be wrapped up into fullerene, rolled into CNT and stacked into graphite.
The successful fabrication of graphene spurred the everlasting pursuit of 2D carbon allotropes,
which can be attained either by meticulous arrangements of non-hexagonal $\mathrm{sp^2}$ defects
within hexagonal framework (or not),
or by bold introduction of non-$\mathrm{sp^2}$ hybridization.
Haeckelite~\cite{Terrones_haeckelite}, biphenylene~\cite{Hudspeth_biphenylene}, T-graphene~\cite{Wang_T-graphene}
and penta-graphene~\cite{Zhang_pentagraphene} fall into first category,
and are periodic and regular arrangement of common defects in graphene~\cite{Banhart_defects},
such as tetrarings, pentagons, hexagons, heptagons, and octagons;
while radialenes~\cite{Dantas_radialenes}, graphyne and graphdiyne family~\cite{Li_graphdiyne,Neiss_graphyne}
belong to second category and contain acetylene ($\mathrm{sp}$ hybridization) or diacetylene ($\mathrm{sp^2}$ hybridization) moieties.
Most of aforementioned 2D carbon allotropes are planar and of monoatomic thickness
due to linear form of acetylene and diacetylene moieties
and planar geometry of paired pentagon-heptagon and pentagon-octagon.

Though various three dimensional $\mathrm{sp^3}$ carbon allotropes were experimentally synthesized
~\cite{Sheng_Tcarbon,Zhang_TcarbonNW,Xu_Tcarbon,bccC8,Johnston_bccC8,Liu_bccC8},
no stable, fully $\mathrm{sp^3}$ hybridized, two dimensional carbon structure has been experimentally synthesized yet.
The difficulty lies in the fact of pyramidization of $\mathrm{sp^3}$ hybridization and hardness to form planar geometry.
Though a theoretical proposal suggested the possibility
to form free-standing and pure $\mathrm{sp^3}$ bonded carbon polymorph,
it suffers phase-transition to other structures~\cite{Sen_cyclobutane}.
Recently, two carbon phases with $\mathrm{sp^2}$-$\mathrm{sp^3}$ networks were theoretically proposed~\cite{Sen_cyclobutane,Wang_hpc18}.
Given the versatile electronic properties and diverse bonding forms of carbon atoms,
the diagram of $\mathrm{sp^2}$-$\mathrm{sp^3}$ carbon phase is still a virgin land,
and numerous structures should exist.
Using first-principles, we explore this phase diagram and find a new 2D carbon allotrope, termed as clathrate graphene,
which is in minimum egg-tray-shape and aesthetically analogous to clathrate patters in Chinese traditional wooden windows.
Our calculations show that the proposed clathrate graphene is dynamically and elastically stable,
and is a semiconducting metastable allotrope with an indirect band gap of 0.9 eV.
Besides, we observe multiple von Hove singularities near the Fermi energy and giant strain tuning effect on the electronic properties.

\section{Theoretical methods}

$\textit{Ab initio}$ package VASP~\cite{Kresse_vasp1} within the projector augmented wave formalism~\cite{Blochl_paw}
is employed to carry out density functional theory (DFT) calculations to optimize atomic structures
and to determine the functional properties of the clathrate graphene.
Perdewe-Burkee-Ernzerhof (PBE) generalized GGA functional~\cite{Marti_pbe}
is adopted to describe the electron exchange and correlation effect.
A plane-wave energy cutoff of 550 eV is utilized to expand the wave function.
Periodic boundary condition and slab geometry are adopted
and vacuum space of appropriate lengths along the non-periodic z direction is added
to avoid spurious interactions between periodic images.
The in-plane cell parameters and atomic positions are fully relaxed
until the force acting on each carbon atom is less then 0.01 eV/\AA.
For structural optimizations and electronic/mechanic property calculations,
Brillouin zone sampling is done with $\Gamma$-centered $15 \times 15 \times 1$ k-point mesh.
Phonon properties are computed using VASP interfaced with Phonopy~\cite{Togo_phonopy} code
using $4 \times 4 \times 1$ supercells and $\Gamma$-centered $3 \times 3 \times 1$ k-point mesh.

\section{Results and discussions}

\subsection{Geometry parameters and energetics}
\begin{figure}[b]
  \centering
  \includegraphics[width=0.4\textwidth]{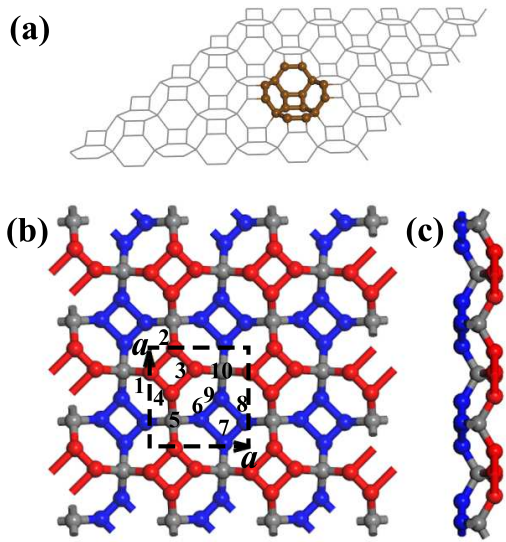}\\
  \caption{(Color online) Illustration of atomic structure of tetragonal clathrate graphene: 
  (a) perspective view, (b) top view, and (c) side view.
  The ball and stick in (a) indicate the minimum egg-tray included in the structure.
  Red and blue color in (b)-(c) denotes the carbon atoms in upper and lower tetrarings respectively.
  }\label{clathrate}
\end{figure}

In Fig.~\ref{clathrate}, we show the atomic structure of proposed carbon allotrope in perspective-, top- and side-view.
The atomic structure, highly resembling the beautiful clathrate patterns in traditional Chinese wooden windows
and possessing an elegant and aesthetic view,
is self-explanatory to its name 'clathrate graphene'.
The structure has the symmetry of $P$-$42_1m$ (IT \#113) and in-plane lattice constants of $a$=$b$=4.68 \AA,
and contains ten carbon atoms in primitive cell.
Two carbon atoms (grey C5 and C10) are $\mathrm{sp^3}$ hybridized and
occupy the crystallographic 2b Wyckoff position,
and eight (red C1-C4 and blue C6-C9) are $\mathrm{sp^2}$ hybridized and
occupy the equivalent crystallographic 8f Wyckoff position.
Due to pyramidization of $\mathrm{sp^3}$ hybridization, clathrate graphene is not monoatomic thick,
alike other 2D $\mathrm{sp^2}$-$\mathrm{sp^3}$ carbon allotropes~\cite{Wang_hpc18,Sen_cyclobutane}.
Two tetrarings formed by red C1-C4 and blue C6-C9 respectively are located on opposite sides of 2b carbon atoms
(see side-view in Fig.~\ref{clathrate}(c)),
and four hexagons formed by $\mathrm{sp^2}$ and $\mathrm{sp^3}$ carbon atoms are bridging these tetrarings.
The tetrarings and hexagons are symmetrically equivalent to each other respectively.
It's noteworthy that the two tetrarings have orthogonal orientations:
the long side of upper tetraring 1234 (C1-C2) is along [110] direction
while the long side of lower tetraring 6789 (C6-C7) is along [$\bar{1}$10].
Though high energy cost of carbon tetraring,
it prevails in many carbon allotropes,
such as 2D T-graphene~\cite{Wang_T-graphene} and 3D bct $\mathrm{C_4}$~\cite{Umemoto_bctC4},
bcc $\mathrm{C_8}$~\cite{bccC8,Johnston_bccC8,Liu_bccC8}, and bct $\mathrm{C_{16}}$~\cite{Cheng_BCTC16},
where bcc $\mathrm{C_8}$ can be prepared
by plasma deposition or pulsed-laser induced liquid-solid interface reaction.
Besides the clathrate pattern, we notice that it is of minimum egg-tray shape
as shown in the perspective view in Fig.~\ref{clathrate}(a).
Recently, egg-tray-shaped graphenes were proposed by Liu et al.~\cite{Liu_eggtray}.
Combining inwards-curved pentagon-hexagon fragments and outwards-curved octagon-hexagon fragments,
they obtained various egg-tray-shaped graphenes by different choice of joint methods and different ratios of hexagons.
However, their egg-tray-shaped graphenes are $\mathrm{sp^2}$ hybridized and much more larger
where the smallest one having 28 carbon atoms and being nearly 3 times larger than our clathrate graphene.
The minimum egg-tray present in clathrate graphene would supply
great absorption site with large coverage density for adatoms,
and will make it highly applicable in various fields,
such as hydrogen storage, anode materials, 2D spintronics~\cite{Liu_eggtray,absorption}.

\begin{table}[t]
\centering
\caption{ Geometry parameters of the tetragonal clathrate graphene: inequivalent bonds and angles.}
\begin{tabular}{cccccccccccc}
\hline\hline
   & bond    &      &      &  bond length (\AA)      &     &     & angle &  &  & angle degree(\degree) &                      \\ \hline
   & 1-2     &      &      & 1.572                   &     &     & $\angle 123$   &  &  & 90                    &             \\
   & 1-4     &      &      & 1.357                   &     &     & $\angle 345$   &  &  & 125.03                &             \\
   & 4-5     &      &      & 1.514                   &     &     & $\angle 456$   &  &  & 104.76                &             \\
   &         &      &      &                         &     &     & $\angle 569$   &  &  & 130.16                &             \\
\hline\hline
\end{tabular}
\label{geo}
\end{table}

As pointed out by Ref.~\cite{Liu_eggtray},
the percentage and deformation of hexagons have large impact on the energetics.
In our clathrate graphene, hexagons have the percentage of 66.7\%.
Besides, the planar geometries of both tetrarings and hexagons are kept
and can be proved by small deviation (on the order of $0.1 \degree$) of summation of all interior angles
away from their ideal values ($360 \degree$ and $720 \degree$).
In Table.~\ref{geo}, we show the inequivalent bonds and angles.
The bonds in rectangular tetraring deviate by large amount from C-C bonds in graphene (1.42 \AA),
its long side length being 1.572 \AA \ and short side length being 1.357 \AA.
In hexagons, four bonds (C4-C5 and its equivalents) connecting $\mathrm{sp^2}$ carbon and $\mathrm{sp^3}$ carbon
have same bond lengths of 1.514 \AA,
which deviate by small amount from C-C bonds in diamond (1.54 \AA).
However, the hexagons contain both the short (C6-C9) and long (C3-C4) bond in tetrarings,
which render the asymmetric angle values.
Besides, the hexagons are prolonged along [110] or [$\mathrm{\bar{1}10}$] direction.
Overall, three different angles exist in hexagons, that is,
104.76 \degree ($\angle 456$ for example),
125.03 \degree ($\angle 345$ for example),
and  130.16 \degree ($\angle 569$ for example).
The calculated energy cost of clathrate graphene with respect to pristine graphene is $\sim 1.1$ eV/Carbon,
higher than pure $\mathrm{sp^2}$ T-graphene ($\sim 0.85$ eV/Carbon)~\cite{Wang_T-graphene},
penta-graphene ($\sim 0.95$ eV/Carbon)~\cite{Zhang_pentagraphene}
and $\mathrm{sp^2}$-$\mathrm{sp^3}$ hP-C18 ($\sim 0.6$ eV/Carbon)~\cite{Wang_hpc18},
but lower than ladderane-derived $\mathrm{sp^2}$-$\mathrm{sp^3}$ carbon allotropes ($\sim 1.3$ eV/Carbon)~\cite{Sen_cyclobutane}.

\subsection{Dynamic stability}

To confirm the dynamic stability of clathrate graphene,
we then calculate its phonon spectrum and show the result in Fig.~\ref{band}(a).
Absence of imaginary frequency in the phonon band structure
demonstrates the dynamic stability of this allotrope.
Due to the two dimensional nature of this structure,
the low-lying acoustic mode (that is, out-of-plane ZA mode) displays a characteristically quadratic dependence on $q$
when approaching $\Gamma$ point~\cite{ZA_mode},
while the longitudinal acoustic (LA) mode and transverse acoustic (TA) mode linearly depend on $q$.
For the high energy optical phonon modes,
a remarkable phonon gap from 36 THz to 46 THz is observed.
Such a large phonon gap would have large restriction effect on the three phonon-phonon scattering phase space
and will have large effect on its thermal properties at high temperature~\cite{phononGap,VA}.

\begin{figure}[b]
  \centering
  \includegraphics[width=0.27\textwidth]{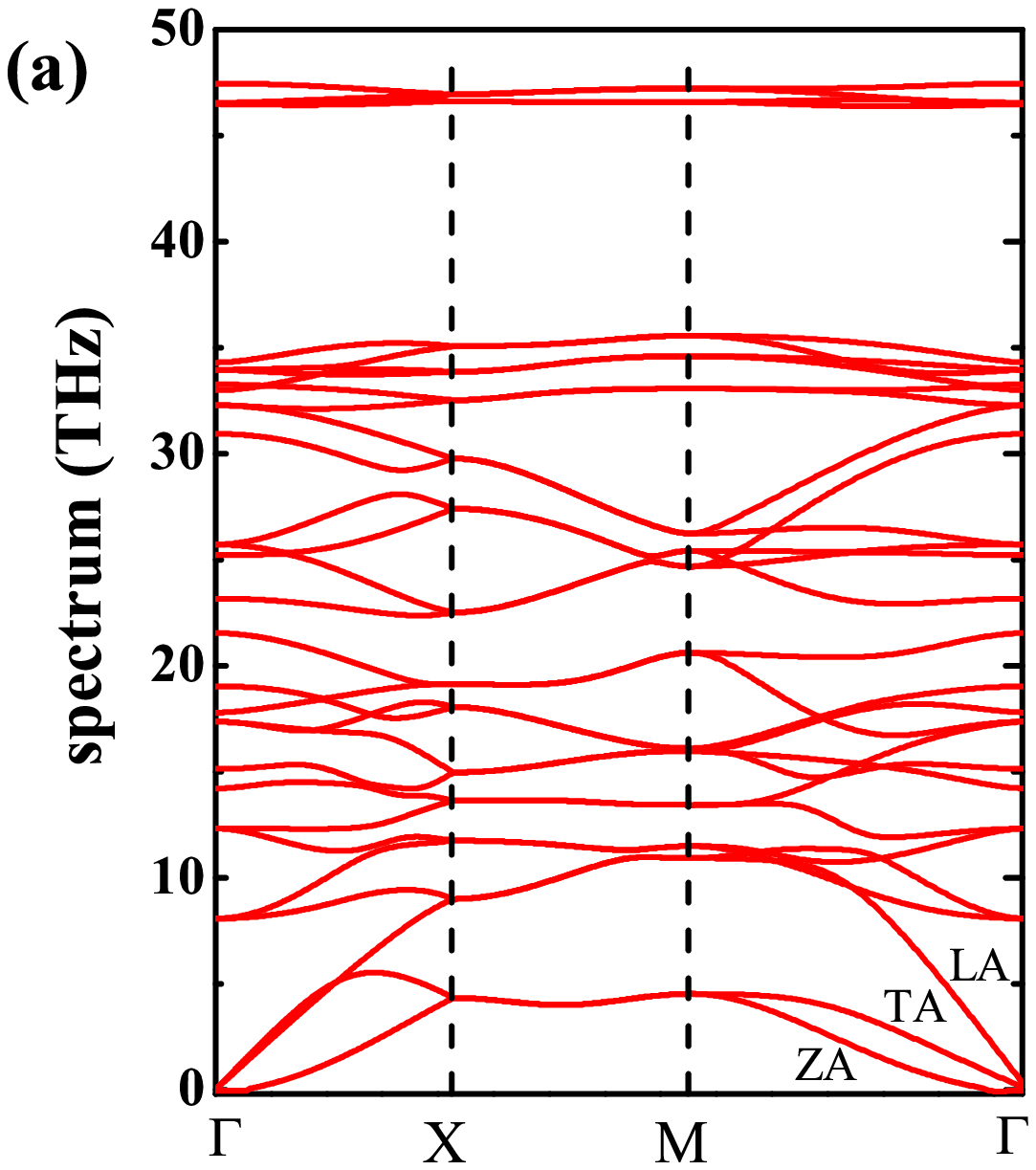}
  \includegraphics[width=0.18\textwidth]{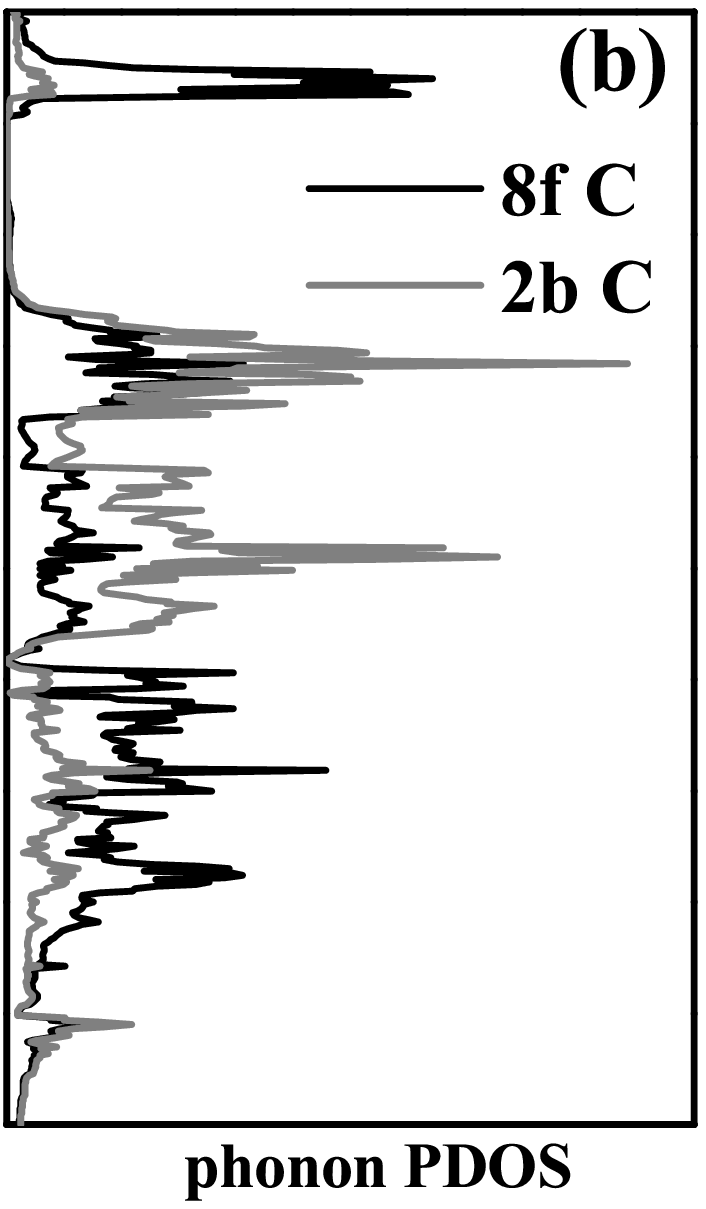}\\
  \caption{(a) Phonon band structures along high symmetry path:
  $\Gamma$(0,0,0)-$\mathrm{X}$(0.5,0,0)-$\mathrm{M}$(0.5,0.5,0)-$\Gamma$(0,0,0),
  and (b) phonon density of states of clathrate graphene.}\label{phonon}
\end{figure}

\begin{figure*}[htb]
  \centering
  \includegraphics[width=0.44\textwidth]{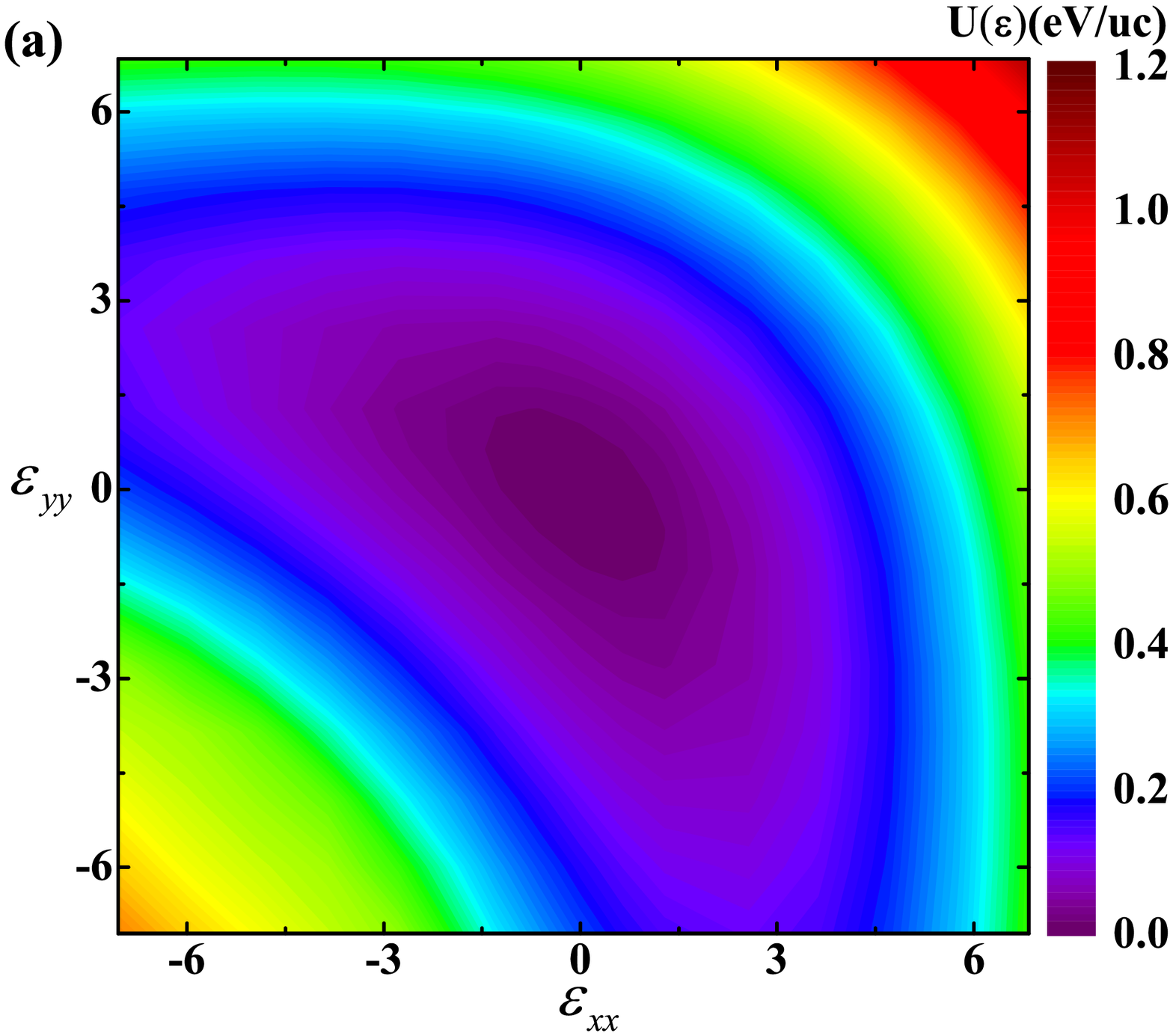}
  \includegraphics[width=0.4125\textwidth]{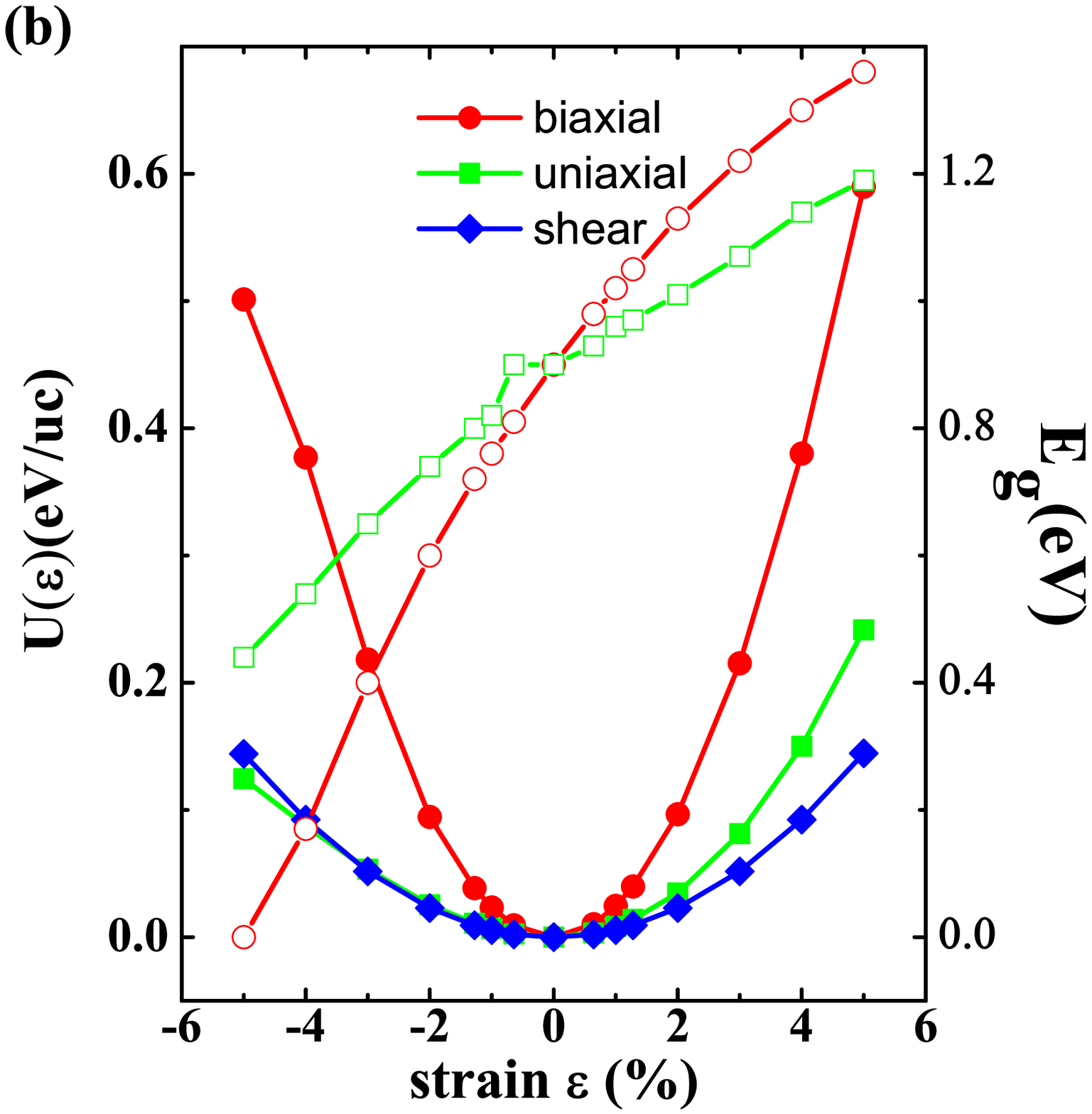}
  \caption{(Color online) (a) Two dimensional contour plot of energy landscape under strains,
  (b) strain energy (left coordinate, solid) and band gap (right coordinate, empty) under various strains.}\label{strain}
\end{figure*}

In Fig.~\ref{band}(b), we show the atom-averaged projected phonon density of states (PDOS) of each class (8f and 2b).
Distinct distribution behaviors are observed for 8f and 2b atoms:
in low energy regime (below 10 THz), each atom of these two classes equally contributes to total DOS;
in medium energy regime (between 10 THz and 20 THz), 8f carbon atoms dominate the contributions;
in medium-high energy regime (30 - 35 THz), 2b carbon atoms dominate;
for highest phonon modes in narrow frequency windows around 46 THz,
the major contribution mostly comes from the 8f carbons.
These behaviors are quite similar to the previous reported three dimensional
$\mathrm{sp^2}$-$\mathrm{sp^3}$ interlocking carbon structure~\cite{interlocking}.

\subsection{Elastic stability and mechanical properties}

In order to ensure the crystal stability,
another criterion besides the previous dynamic stability criterion (no imaginary phonon frequency) must be fulfilled,
that is, Born elastic stability criteria to guarantee the positive-definiteness of strain energy~\cite{Born}.
For two dimensional crystals,
the energy of the crystal under small strain follows a quadratic form~\cite{Zhang_pentagraphene,Ding_SiX}:
\begin{equation}\label{cij}
\begin{split}
  U(\epsilon)& =E(\epsilon)-E_0 \\
  & = \frac{1}{2}C_{11}\epsilon_{xx}^2 + \frac{1}{2}C_{22}\epsilon_{yy}^2 + C_{12}\epsilon_{xx}\epsilon_{yy}
  +2C_{66}\epsilon_{xy}^2
\end{split}
\end{equation}
where $E(\epsilon)$ and $E_0$ are the total energies of the stressed and unstressed system respectively,
$\epsilon$ is the strain and $C_{ij}$ is element of elastic constants matrix.
In this expression, standard Voigt notation is used for the indices of strain,
that is, 1-xx, 2-yy, and 6-xy.
For 2D cases, the elastic stability conditions~\cite{Zhang_pentagraphene} are
\begin{equation}\label{BornCondition}
  C_{11}C_{22}-C^2_{12} > 0, C_{66}>0
\end{equation}
Since the tetragonal symmetry of clathrate graphene, $C_{11}=C_{22}$,
above conditions are simplified as
\begin{equation}\label{TetraBornCondition}
  C_{11}> |C_{12}|, C_{66}>0
\end{equation}

We show two dimensional contour plot of the energies (with respect to ground unstressed state)
of stressed clathrate graphene in Fig.~\ref{strain}(a),
the strain energies (left coordinate, solid) under the biaxial, uniaxial and shear strain in Fig.~\ref{strain}(b).
Manifested anisotropic features are observed for strain energy in the contour plot.
In Fig.~\ref{strain}(b),
asymmetry of strain energy becomes notable when $|\epsilon|> 2.0\%$.
In small strain regime ($|\epsilon|< 2.0\%$),
the strain energy is symmetric and thus Eq.~\ref{cij} is applicable,
thus $U(\epsilon)=1/2C_{11} \epsilon_{xx}^2$ for uniaxial strain $\epsilon_{xx} \neq 0, \epsilon_{yy}=0$,
$U(\epsilon)=(C_{11}+C_{12}) \epsilon_{xx}^2$ for biaxial strain $\epsilon_{xx}=\epsilon_{yy}$.
By parabolic fitting these curves,
we obtain $C_{11}=$ 110 N/m, $C_{12}=$ 64 N/m, $C_{66}=$ 21 N/m,
satisfying the Born stability condition and validating its mechanical stability.
After obtaining these elastic coefficients,
the in-plane Young's modulus can be evaluated using following equation
\begin{equation}\label{Young}
  E=\frac{C_{11}^2-C_{12}^2}{C_{11}}
\end{equation}
then we obtain $E$=73 N/m.
The calculated $C_{11}$ and $E$ of clathrate graphene are
about one-third of pristine $\mathrm{sp^2}$ graphene (354 N/m and 345 N/m)~\cite{Lee}
and $\mathrm{sp^2}$-$\mathrm{sp^3}$ hp-C18 (369 N/m and 349 N/m)~\cite{Wang_hpc18},
, and are half of penta-graphene (265 N/m and 264 N/m)~\cite{Zhang_pentagraphene},
and are comparable with various $\mathrm{sp^2}$ egg-tray shaped graphenes~\cite{Liu_eggtray}.

\subsection{Electronic properties}

\begin{figure}[htb]
  \centering
  \includegraphics[width=0.27\textwidth]{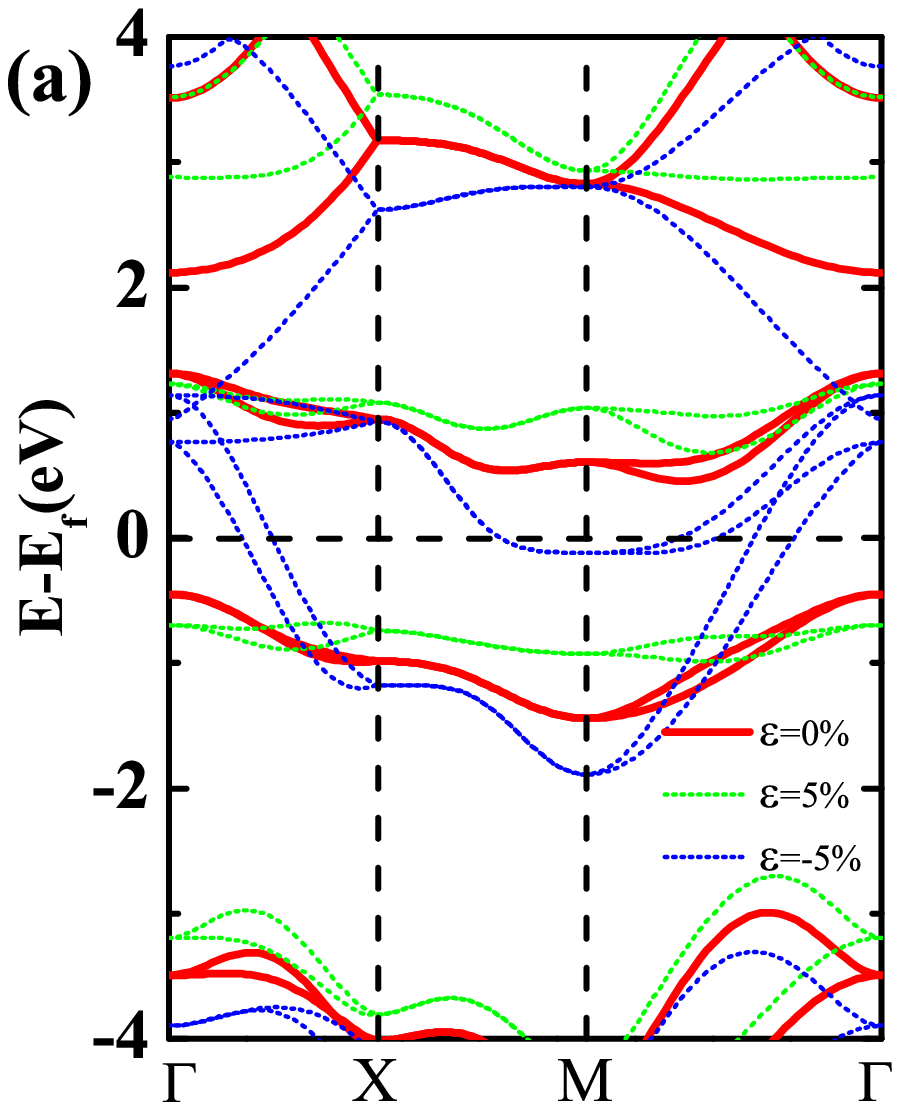}
  \includegraphics[width=0.18\textwidth]{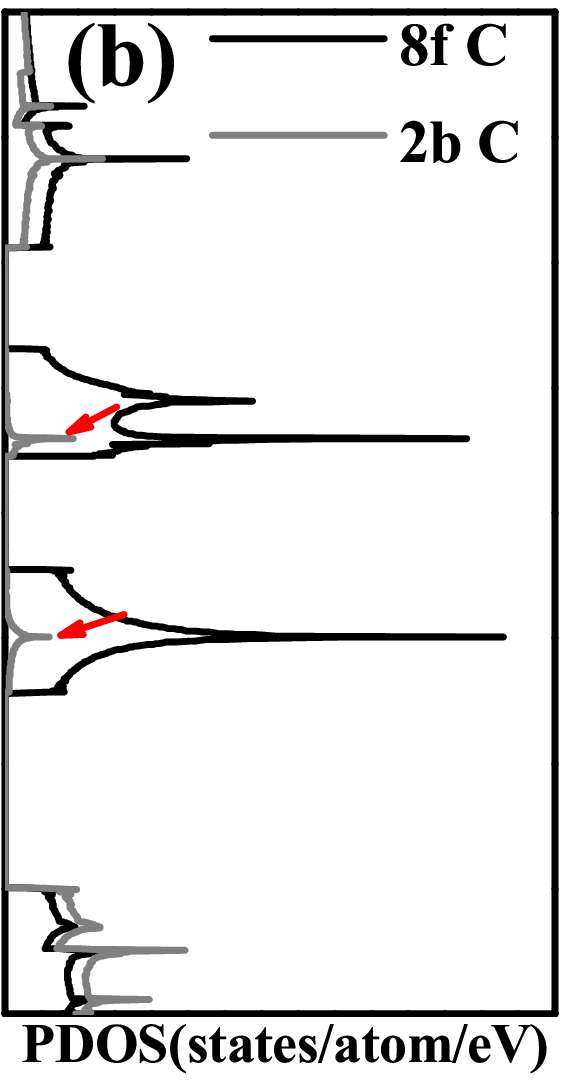}\\
  \caption{(a) Electronic band structures along high symmetry path under biaxial strain of 0\% (red solid),
  5\% (green dotted), and -5\% (blue dotted) respectively. Fermi energy is set in the middle of CBM and VBM for semiconducting states.
  The path is set as $\Gamma$(0,0,0)-$\mathrm{X}$(0.5,0,0)-$\mathrm{M}$(0.5,0.5,0)-$\Gamma$(0,0,0).
  (b) Atom-averaged projected density of states (PDOS) of unstressed clathrate graphene.}\label{band}
\end{figure}

We also investigate the electronic properties of this allotrope.
The band structure along high symmetry directions and atom-averaged projected density of states (DOS)
are shown in Fig.~\ref{band}(a) and (b) respectively.
For unstressed clathrate graphene (red solid),
both the highest two valence bands and the lowest two conduction bands degenerate
in the Brillouin zone around $\Gamma$ point and in k-path from $\mathrm{X}$ to $\mathrm{M}$.
The structure is revealed as an indirect semiconductor
with its valence band maximum (VBM) located at $\Gamma$ point
and conduction band minimum (CBM) located between $\mathrm{M}$ and $\Gamma$ point,
The band gap is 0.9 eV in the PBE level,
half of hp-C18 (indirect band gap of 1.86 eV )~\cite{Wang_hpc18}.
DOS projection analyses show that
majority contributions of both conduction bands and valence bands near the Fermi energy
come from the $\mathrm{sp^2}$ 8f carbon atoms (C1-C4 and C6-C9).
Besides, we observe two obvious von Hove singularities (VHS) close to Fermi level in conduction and valence bands,
and indicate them by red arrows in Fig.~\ref{band}(b).
At these VHS, DOS of both 8f and 2b carbon atoms soar up,
indicating the strong hybridization of 8f and 2b carbon atoms.
In pristine graphene, VHS occurs at energies far away from Fermi energy~\cite{Castro_graphene},
the VHS close to Fermi level presented in clathrate graphene may indicate new interesting phenomenon~\cite{Kohn_VHS,Rice_VHS}.

Strain tuning is an effective method to manipulate the electronic properties of two dimensional materials.
To this end, we calculate the evolution of band gap ($E_g$) under in-plane biaxial and uniaxial strains
within range of [-0.05, 0.05] and show the results in Fig.~\ref{strain}(b) (right coordinate, empty).
For both uniaxial and biaxial strain within the considered range,
the band gaps monotonously increase when the strains change from compressive to tensile.
However, the biaxial strain is more effective:
the compressive strain of 5\% would make the sample metallic.
In order to probe the origin of strain modulation effect on band gap,
we plot in Fig.~\ref{band} the band structure
with biaxial tensile 5\% (green dotted) and compressive -5\% strain (blue dotted).
Under tensile strain,
both the lowest two conduction bands and the highest two valence bands
move down around $\Gamma$ point and move up in other k points.
Accordingly, CBM remains between $\Gamma$ and $\mathrm{M}$ point and moves up,
VBM changes from $\Gamma$ point to a $k$ point between $\Gamma$ and $\mathrm{X}$ and moves down.
Overall, the $E_g$ is enlarged for tensile strain.
For compressive strain of -5\%,
the degeneracy of valence bands and conduction bands in unstressed case is eliminated.
Besides, these bands shift by large amount and in opposite direction:
the whole lowest two conduction bands move down
and the two highest valence bands around $\Gamma$ point move up,
they cross at $\Gamma$ point and another point between $\Gamma$ and $\mathrm{X}$
and change the semiconducting nature to metallic.
The giant biaxial strain tuning effect on the electronic properties of clathrate graphene
would have potential applications in carbon-based nanoelectronics.

\section{Conclusions}

In summary, using \textit{ab initio} calculations,
we explore the phase diagram of two dimensional carbon allotrope
with $\mathrm{sp^2}$-$\mathrm{sp^3}$ network,
and propose a new clathrate-like carbon allotrope with minimum egg-tray shape, termed as clathrate graphene.
The tetragonal clathrate graphene contains 8 $\mathrm{sp^2}$ carbon atoms
and 2 $\mathrm{sp^3}$ carbon atoms in its primitive unit cell,
and two perpendicularly oriented rectangles and four bridging hexagons.
Both rectangles and hexagons retain their planar geometry though deformed bond lengths and bond angles.
Calculations demonstrate that the clathrate graphene is metastable with 1.1 eV/carbon higher than graphene,
and is energetically comparable with pure $\mathrm{sp^2}$ $T$-graphene and penta-graphene.
Further calculations show that this clathrate graphene is dynamically and elastically stable,
and is semiconductor with an indirect band-gap of 0.90 eV.
Biaxial compressive strain of 5\% is found to turn the structure metallic,
and the he giant strain tuning effect is attributed to the strain-induced degeneracy breaking and significant band shift.
The findings are expected to have potential nanoelectronic applications and help search new carbon allotropes.

\section{Acknowledgements}
G.H. Zheng acknowledges funding from NSF of China
(Grants No. 11847049) and from Guizhou University (Grants No. 2017-66).
X.S. Qi acknowledges the financial support from Platform of Science and
Technology and Talent Team Plan of Guizhou Province
(Grant No. 2017-5610)


\end{document}